\documentclass[12pt]{article}

\usepackage{graphics}
\usepackage{epsfig}
\usepackage{a4wide}

\newcommand{\be}{\begin{equation}}
\newcommand{\ee}{\end{equation}}
\newcommand{\bea}{\begin{eqnarray}}
\newcommand{\eea}{\end{eqnarray}}
\newcommand{\bfig}{\begin{figure}}
\newcommand{\efig}{\end{figure}}
\newcommand{\bc}{\begin{center}}
\newcommand{\ec}{\end{center}}
\newcommand{\btab}{\begin{tabular}}
\newcommand{\etab}{\end{tabular}}

\newcommand{\szz}{\sigma_{zz}}
\newcommand{\sxx}{\sigma_{xx}}

\newcommand{\sxz}{\sigma_{xz}}
\newcommand{\srr}{\sigma_{rr}}

\newcommand{\srz}{\sigma_{rz}}
\newcommand{\stt}{\sigma_{\theta\theta}}
\newcommand{\sij}{\sigma_{ij}}

\newcommand{\uz}{u_z}

\newcommand{\ur}{u_r}

\newcommand{\dr}{\partial}

\newcommand{\osl}{\textsc{osl}}

\newcommand{\De}{\Delta}

\newcommand{\intq}{\int_0^{+\infty} \!\!\!\!\!\!\! dq \,}

\begin{document}

\title{Stress response function of a granular layer:\\
quantitative comparison between experiments\\
and isotropic elasticity}
\author{Dan Serero, Guillaume Reydellet, Philippe Claudin
and \'Eric Cl\'ement\\
Laboratoire des Milieux D\'esordonn\'es et H\'et\'erog\`enes
(UMR 7603 du CNRS)\\
4 place Jussieu - case 86, 75252 Paris Cedex 05, France.\\
\\
Dov Levine\\
Technion - Israel Institute of Technology\\
Physics department, 32000 Haifa, Israel.}
\date{\today}
\maketitle

\begin{abstract}
We measured the vertical pressure response function of a layer of sand
submitted to a localized normal force at its surface. We found that this
response profile depends on the way the layer has been prepared: all
profiles show a single centered peak whose width scales with the thickness
of the layer, but a dense packing gives a wider peak than a loose one. We
calculate the prediction of isotropic elastic theory in presence of a bottom
boundary and compare it to the data. We found that the theory gives the
right scaling and the correct qualitative shape, but fails to really fit the
data.
\end{abstract}

\vspace{2cm}

\noindent
\underline{PACS numbers}:\\
\begin{tabular}{ll}
46.25.-y & Static elasticity \\ 
45.70.Cc & Static sandpiles \\ 
83.70.Fn & Granular solids
\end{tabular}

\newpage

\section{Introduction}
\label{intro}

The statics of granular materials has been receiving recently a lot of
attention, for a review see e.g. \cite{deGennes}. An important issue is still
to understand the mechanical status of an assembly of non-cohesive grains. In
the small deformation limit, a classical viewpoint assumes a behavior akin to an
effective elastic medium. At a given confinement pressure, linear relations
between stress and strain are measured and for larger strains,
another picture is proposed based on a plastic modelling of the
stress-strain relations. Therefore, for all practical purposes the available
models used to describe granular matter in the quasi-static limit are of the
elasto-plastic class with constitutive parameters determined empirically
from standard triaxial tests \cite{Wood}. This elastic viewpoint is somehow
corroborated by ultrasound propagation experiments where, under large
confining pressure, elastic moduli of $p$ and $s$ waves produced by a
localized pulse can be measured \cite{Jia}. However, sound propagation
measurements also evidence a strong `speckle-like' component associated
with the intricate contact force-paths topology or `force chains' network.
In a granular packing, contact forces of amplitude larger than the average
were found to organize in cells of sizes of about $10$ grains diameters
\cite{Radjai}. The fragile character of these structures is even more obvious
at low confining pressure where, for example, ultra-small perturbations within
the pile can completely modify the sound response spectrum \cite{Liu}. More
generally, subtle self-organization properties of the contact force network
(also called the texture) were evidenced by thorough numerical studies by
Radjai et al. \cite{Radjai}.

At a macroscopic level, the pressure profile under the base of a sand heap
built from a from a point source (i.e. from a hopper outlet), shows a minimum
below the apex , but \emph{does not} when the heap is constructed by
successive horizontal layers \cite{Smid,Vaneltas}. This surprising effect is
currently viewed as a signature of the preparation history. Successive
avalanches originated from the hopper outlet could have embedded a microscopic
structure which is reflected macroscopically by an arching effect below the
apex.

The ability for a granular piling to change its texture (granular contact
network, force chains geometry) in response to an external constraint, have
cast legitimum suspicions on the fundamental validity of elasticity for packing
of hard grains. For these reasons, a new class of models -- called \osl\ for
`Oriented Stress Linearity' -- was introduced by Bouchaud et al.
\cite{bcc}, which could explain remarkably well the sandpile data \cite{wcc},
as well as stress screening in silos \cite{Vanelsilo}. These models have been
the subject of a rather controversial debate \cite{Savage}. One of the
reasons for that was the fact that they do not belong to the standard
elasto-plastic class. As a matter of fact, they do not require the introduction
of a displacement field, and the usual stress-strain relations are rather
replaced by `stress-only' ones which encode the history dependent state
of equilibrium of the piling. In particular, the equations governing the
stress distribution in these models are of hyperbolic type, which contrasts
with the elliptic (or mixed elliptico-hyperbolic) equations of elastic (or
elasto-plastic) modellings. An attractive feature of hyperbolic equations is
that they have characteristic lines along which stress is transmitted, and
which were argued to be the mathematical transcription of force chains that
one can clearly see in granular systems \cite{Dantu}.

Measurements of the pressure response of a layer of sand submitted to a
localized normal force at its surface soon appeared to be a way to
discriminate between the different classes of models \cite{deGennes}. Such a
crucial experiment addresses at the deepest level, questions on the real
mechanical status for a granular assembly. In elasticity, the shape of this
pressure profile shows a single centered broad peak, whose width scales with
the height $h$ of the layer \cite{Landau}. On the contrary, \osl\ models
predict a response with two peaks (or a ring in three dimensions) on
each sides of the overloaded point. Experiments \cite{manip3D,manip2D,Chicago}
and simulations \cite{Eloy,Moreau} have then been performed recently .
Although the picture is far from being completely clear yet, the conclusions
of these works can be roughly summarized as follows. For disordered systems,
experiments definitively show elastic-like response, while regular packing
exhibit \osl\ features. A third class of granular assemblies have also
macroscopic equilibrium equations of the hyperbolic type: packings that can be
prepared under the special isostaticity condition \cite{Edwards,Tkachenko},
defining the uniqueness of the contact forces once the list of contacts is know
\cite{Roux,Moukarzel}. In pratice, this condition would correspond to a minimum
number of contact per grains, such as frictionless contact forces for 2D random
packing. A recent numerical result obtained for such an assembly
explicitely shows an \osl-like propagation as a response to a localized force
(at least on a scale up to 20 grains size) \cite{Head}. Note at last that, by
contrast, the experiment presented in \cite {daSilva} rather claims a `diffusive'
response function, in agreement with the stochastic scalar $q$-model \cite{qmodel}
but these experiments were performed on a small size packing and in a rather
specific geometry.

In this paper, we present response function measurements obtained on large
pilings made of natural sand. We show that it is possible to get rather
different pressure profiles when preparing the packing with two different
procedures. As in \cite{manip3D}, we found that elastic predictions give the
right scaling and the correct qualitative shape but here, we perform a
quantitative comparison between experimental data and isotropic elasticity
predictions. We seek to answer precisely the question whether stress
transmission properties can still be described using an isotropic elastic
medium theory. What we found is that elasticity actually fails to really fit
the data.

The paper is organized as follows. In section \ref{exp} we expose how the
measures are done and the way the data are calibrated. Then we show how
different data-set can be obtained, depending on the sample preparation
method. Section \ref{elastic} is devoted to the calculation of the stress
components at the bottom of an isotropic elastic layer of finite thickness
$h$. In order to make this paper easier to read, the details of these
calculations are given in appendices \ref{calcul2d} and \ref{calcul3d} for
the two and three dimensional cases respectively. The comparison between the
experiments and elasticity is done in section \ref{fits}. At last, we
conclude the paper with a discussion on the interpretation of the results in
section \ref{conclu}.


\section{The response function experiment}
\label{exp}

The experimental technique that we use for the measure of the pressure
response of a granular layer to the application of a localized vertical
force $F$ at its top surface has been described in details in \cite{manip3D}.
The sketch of the set-up can be seen in figure \ref{manipschema}. Briefly,
the pressure $P$ is measured by the tiny change of the electrical capacity
of the probe due to the slight deformation of its top membrane. $F$ is
applied with a piston whose displacement is monitored and controlled to stay
as small as possible (less than 500$\mu $m). To gain sensitivity, $F$ is
modulated at a frequency $f$, and the probe signal is directed to a lock-in
amplifier synchronized at $f$ too. Any choice of $f$ between $0.1$ and $80$ Hz
gives the same result. We checked that the response $P$ is linear in $F$.
Both piston and probe have a surface in contact with the grains of
$\sim 1$ cm$^{2}$. The container is large enough ($50 \times 50$ cm$^2$) to be
able to neglect finite size effects due to the lateral walls. Its bottom plate
is very rigid (Duraluminium, thickness $2$ cm), and covered by a sheet of sand
paper, in order to avoid sliding of the grains on the plate.

\bfig[t]
\bc
\epsfxsize=8.5cm
\epsfbox{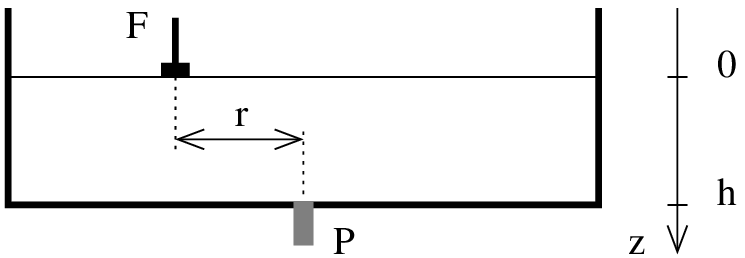}
\caption{\small Sketch of the experimental set-up. A localized vertical
force $F$ is applied on the top surface of the granular layer ($z=0$). The
corresponding pressure response on the bottom is measured at some distance $r$
from that point. The vertical $z$-axis points downwards and we note $h$ the
thickness of the layer (between $0$ and $10$ cm). We use natural `Fontainebleau'
sand whose typical diameter is $\sim 0.3$ mm.
\label{manipschema}}
\ec
\efig

We call $r$ the horizontal distance between the piston and the probe. In
order to measure the profile $P(r)$, it is easier to vary $r$ by moving the
piston. In principle, the horizontal integral of the pressure profiles $%
F^*=\int_0^{+\infty}\!\! dr 2\pi r P(r)$ should be constant and equal to the
force $F$ applied at the surface. In fact, due to arching screening effects
around the probe, this integral actually shows a large dispersion -- see
figure \ref{Fstar} -- from an experiment to another and remains less than
$F$. This screening effect is well known to be inherent to every mesurement of
stresses in granular materials \cite{Savage}, but we could get rid of this
problem of screening we use $F^*$ to renormalize the pressure measurements:
$P \leftarrow \frac{1}{F^*} P$.
The accurate determination of $F^*$ is a
crucial point when it comes to the quantitative comparison between experiments
and theory. In particular, we were very carreful to take enough data points to
have a good estimation of the experimental off-set at large $r$.

\bfig[t]
\bc
\epsfysize=9cm
\epsfbox{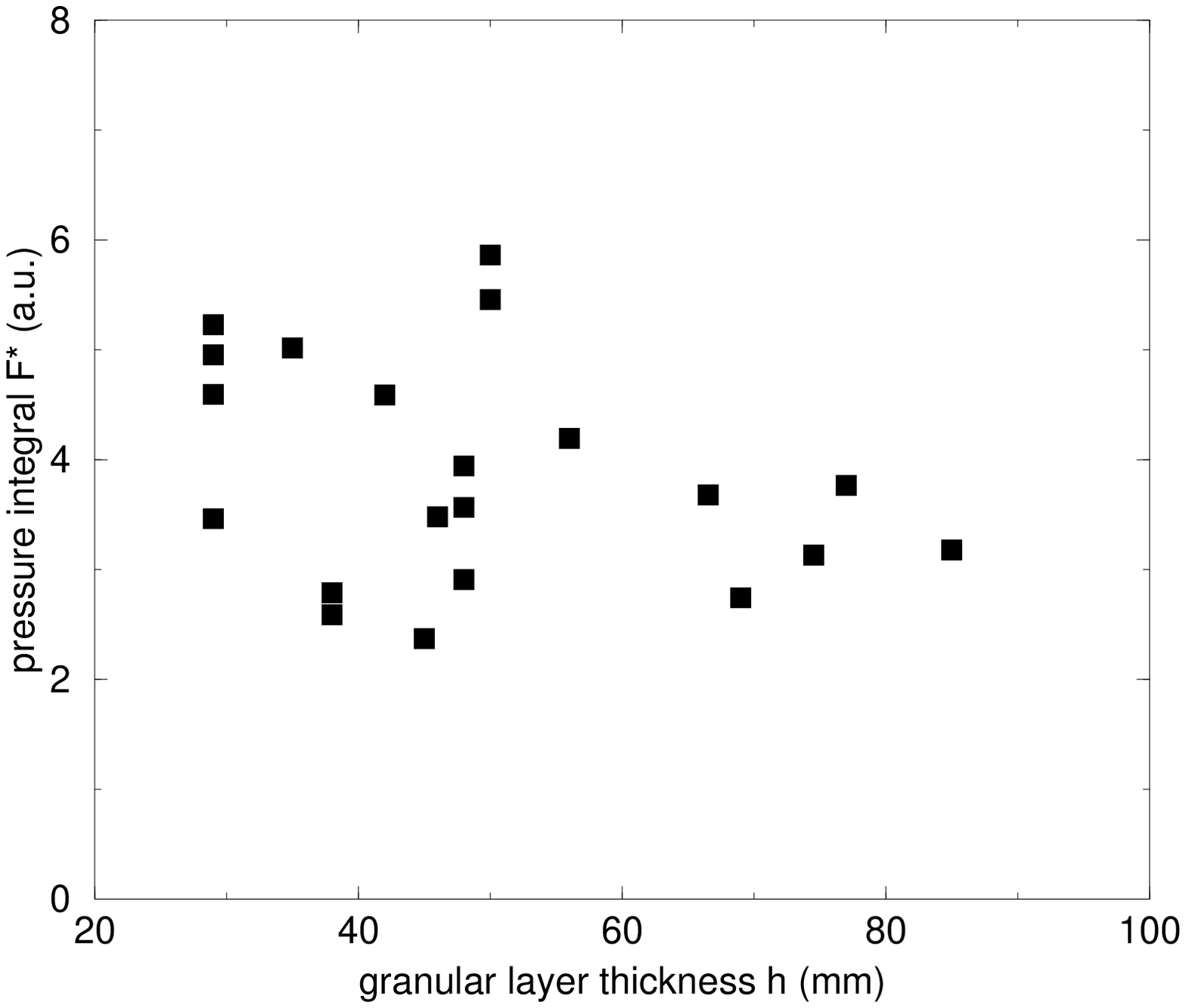}
\caption{\small This plot shows the integral $F^*$ of the profiles $P(r)$ taken
from different experiments. The dispersion is quite large, but no
systematic variation of $F^*$ with the layer thickness $h$ can be evidenced.
\label{Fstar}}
\ec
\efig

\bfig[p]
\bc
\epsfxsize=7.5cm
\epsfbox{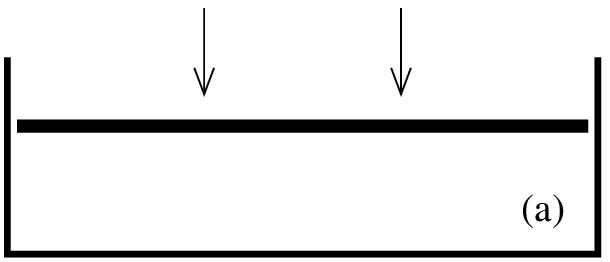}
\epsfxsize=7.5cm
\hfill
\epsfbox{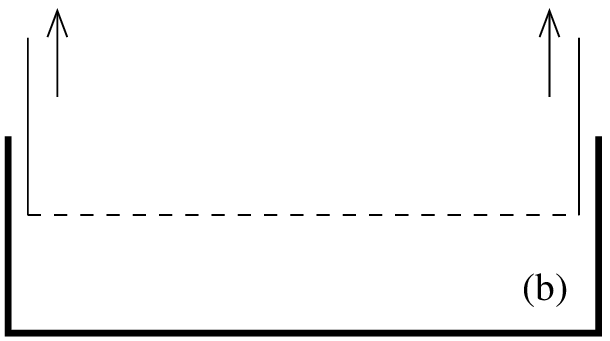}
\caption{\small The packing of the grains has been prepared in two different
ways: we either make it very dense (compacity $\sim 0.7$) by pushing hard on
the grains with a metallic plate (a), or very loose (compacity $\sim 0.6$)
by pulling up a sieve through them (b).
\label{manipprepa}}
\ec
\efig
\bfig[p]
\bc
\epsfysize=9cm
\epsfbox{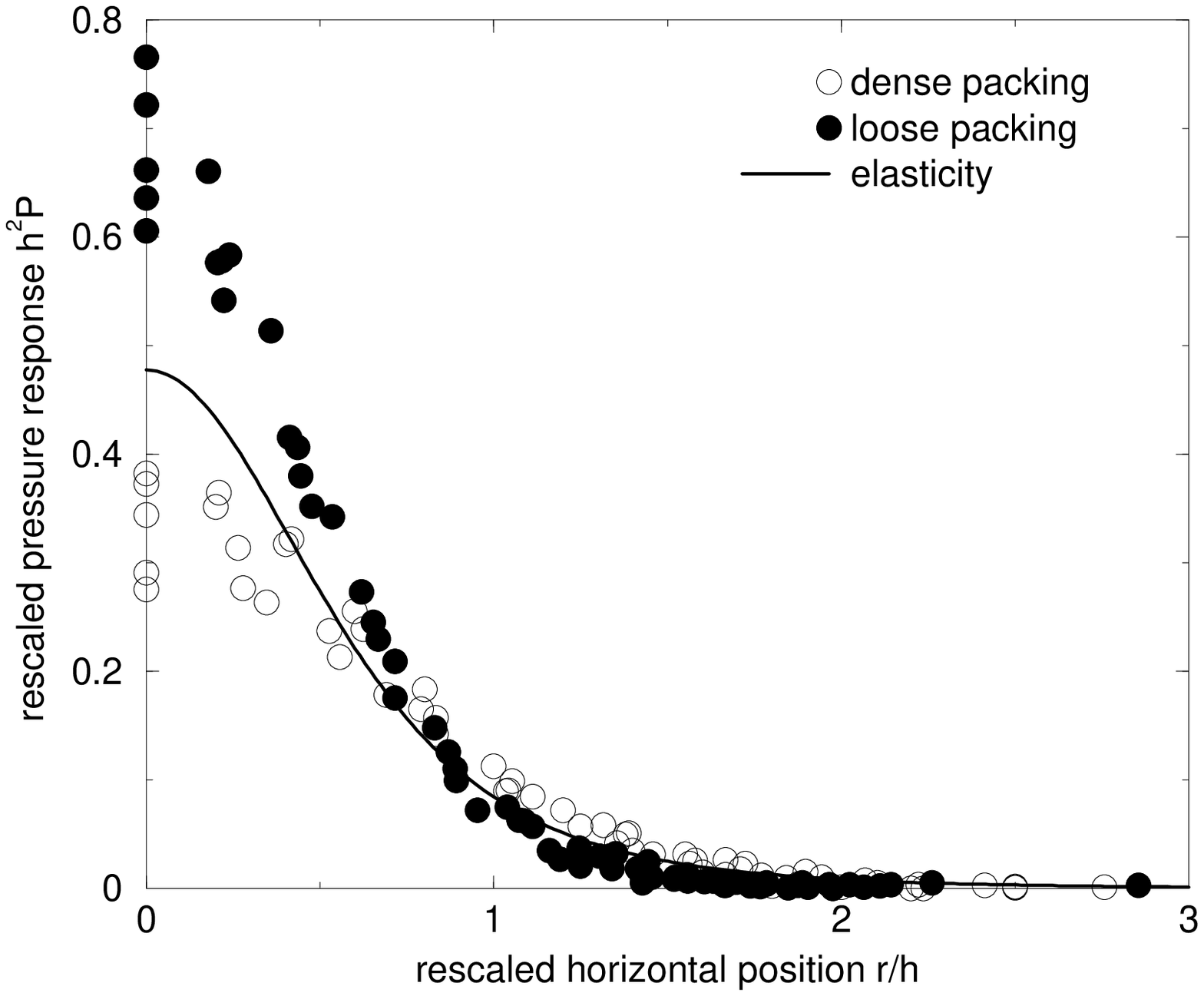}
\caption{\small The pressure response profile depends on the way the system
of grains was prepared: it is broader for a dense packing (empty circles)
than for a loose one (filled circles). The response of a semi-infinite
isotropic elastic medium (solid line) lays in between. For a given
preparation, the experimental data collapse pretty well when renormalized
by their $F^*$ factor -- see text --, and rescaled by the height $h$ of the
layer. Here, we have plotted together measures on layers whose thickness
varies from $h \sim 30$ to $h \sim 60$ mm.
\label{manipresult}}
\ec
\efig

A particular attention should be paid to the way the granular layer has been
prepared. In order to observe different mechanical behavior, we chose two
extreme procedures. There are schematized on figure \ref{manipprepa}. The
first one consists of making a packing as dense as possible. We add the sand
by layers of 0.5 cm and after each layer, we push hard on the grains with a
metallic plate. We can then reach a compacity of order of $0.7$ -- note that
this `layer by layer' procedure may create inhomogeneities in the density
field. By contrast, to make the packing very loose, we first place a sieve
on the bottom plate of the container, pour the grains into the box, and then
gently pull up the sieve all through the grains. The corresponding compacity
is of order of $0.6$.

As already mentioned in \cite{manip3D}, data taken from layers of several
heights can be plotted together by rescaling lengths by $h$ -- which
contradicts the `diffusive' description as proposed by the $q$-model. The
rescaled data $h^2P$ as a function of $r/h$ can be seen on figure
\ref{manipresult}. This plot clearly shows that the response of the granular
layer is `history dependent': the pressure profile of a dense packing is much
broader than that of a loose one.

Bousinesq and Cerruti gave the expression of the stress response in the case
of a isotropic semi-infinite elastic medium submitted to a localized and
vertical unitary force $F$ at $r=0$ \cite{Johnson}. For the vertical pressure at
point $(r,z)$, this expression is 
\begin{equation}
\sigma _{zz}=\frac{3F}{2\pi }\,\frac{z^{3}}{(r^{2}+z^{2})^{5/2}}.
\label{szz3dsemiinf}
\end{equation}
This formula is independent of the Poisson coefficient $\nu $ of the elastic
material and thus does not have any adjustable parameter. It is therefore
unable to reproduce the two different experimental pressure profiles. As a
matter of fact, this function lays in between the two profiles -- see figure 
\ref{manipresult}. The pressure responses of a dense and a loose packing of
sand are thus respectively broader and narrower than the standard elastic
response profile.

In the next section, we shall take into account the finite thickness of the
layer and derive the corresponding expressions for the stresses, which will
indeed depend on $\nu$. These expressions will then be quantitavely compared
to the experimental data.

\section{Elastic calculations}
\label{elastic}

In this section, we derive the expressions of the stress tensor
components at the bottom of an isotropic elastic layer of finite thickness
$h$. This calculation is not new, but as far as we know, the available
litterature only provides numerical tables \cite{Giroud} that make fits
difficult to perform. Such a calculation is a bit heavy, and we chose to
present here its main lines only. The full details can be found in
appendices \ref{calcul2d} and \ref{calcul3d} for the two and three
dimensional cases respectively. The formalism we use and the way the
calculation is lead is directly inspired from \cite{Landau,Muskhelishvili}.

The stress state of an elastic material is described by its stress tensor
components $\sigma_{ij}$. At equilibrium, these quantities must verify the
force balance equations: 
\begin{equation}
\label{equil}
\nabla_i \sigma_{ij} = \rho g_j
\end{equation}
where $\rho$ is the density of the material and $g_j$ the gravity vector.
These relations are not enough to form a closed system of equations. An
additional physical input is required. In plain elasticity theory, a
displacement field $u_i$ is introduced -- it measures the change in
position, with respect to the reference state where no constrains are
applied--, and the corresponding strain tensor $u_{ij}=\frac{1}{2} \left (
\frac{\dr u_i}{\dr x_j} + \frac{\dr u_j}{\dr x_i} \right )$
is related to the stresses via linear relations which involve two
parameters which characterize this pure elastic material: its Young modulus
$Y$ and its Poisson coefficient $\nu$.

It is possible to express all the equations in terms of the stress
components only. Eliminating the $u_{ij}$, one gets 
\begin{equation}
\label{equasLandau}
(1+\nu) \De \sij + 
[1+(3-d)\nu] \frac{\dr^2 \sigma_{kk}}{\dr x_i \dr x_j} = 0,
\end{equation}
where $d$ is the space dimension -- these equations are not valid in the
case of a non-uniform external body force. In particular, contracting $i$
and $j$, we see that the trace of the stress tensor is an harmonic function,
i.e. that $\Delta \sigma_{kk} = 0$. These relations include (derivatives of)
the force balance equations (\ref{equil}). Taking the laplacian of
(\ref{equasLandau}), we also see that the $\sigma_{ij}$ are bi-harmonic.

The solutions of equations (\ref{equasLandau}) can be found in Fourier
transforms. Let first focus on the two-dimensional $(x,z)$ case. Because
$\Delta\Delta\sigma_{ij}=0$, the general form of the vertical pressure
$\sigma_{zz}$ can be written as follows: 
\begin{equation}
\label{szzform}
\szz = \intq \cos(qx) \left \{
\left [A_{zz}^+(q) + qz B_{zz}^+(q) \right ] e^{qz} +
\left [A_{zz}^-(q) + qz B_{zz}^-(q) \right ] e^{-qz}
\right \}.
\end{equation}
The expression for $\sigma_{xx}$ is very similar. For the shear stress
$\sigma_{xz}$, the cosinus factor should be replaced by $\sin(qx)$. In fact,
only four of these twelve functions $A$'s and $B$'s are independent. They
are fully determined by the boundary conditions, and we can get this way
explicit -- but integral -- expressions for the stresses.

\bfig[t]
\bc
\epsfysize=9cm
\epsfbox{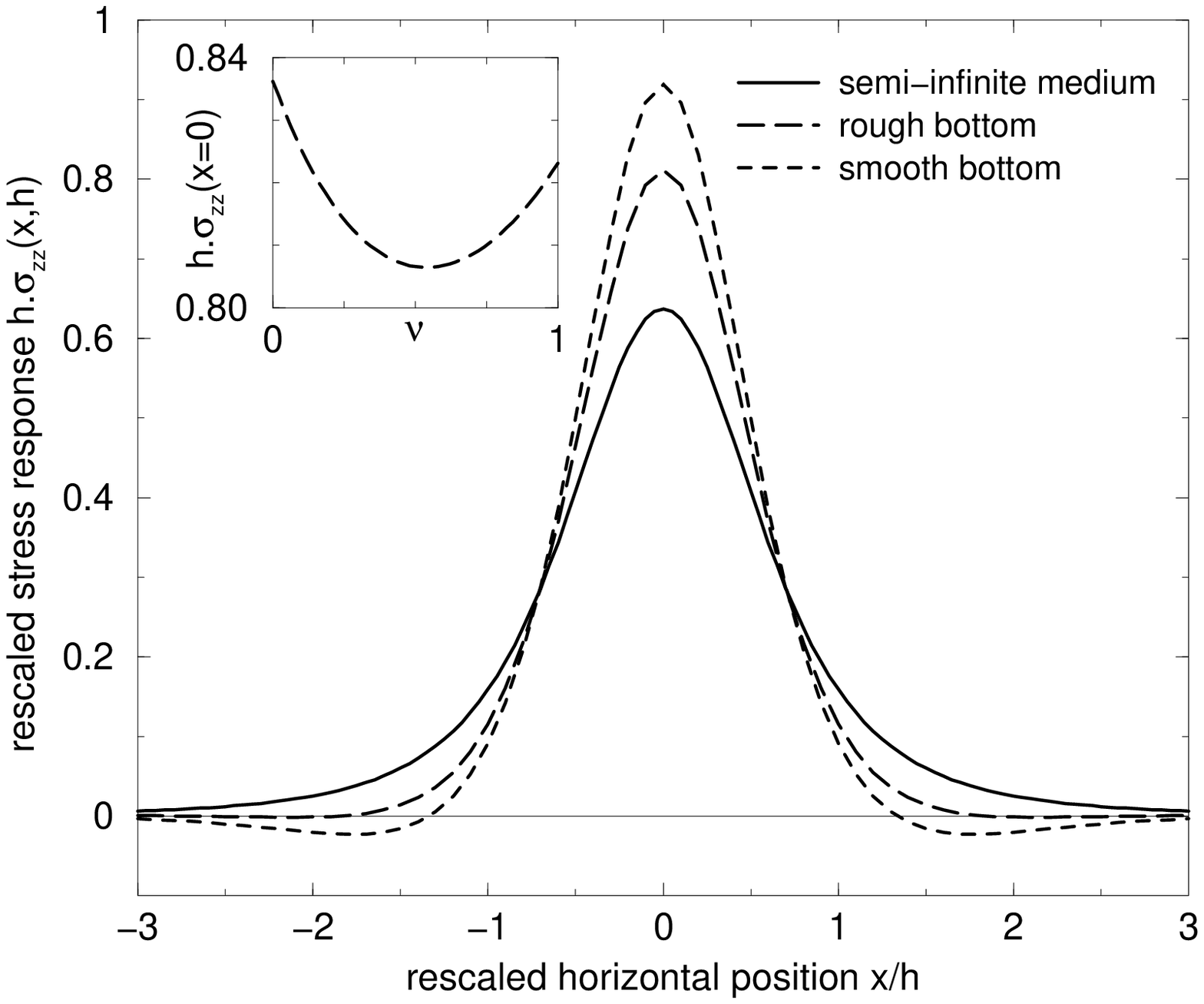}
\hfill
\epsfysize=9cm
\epsfbox{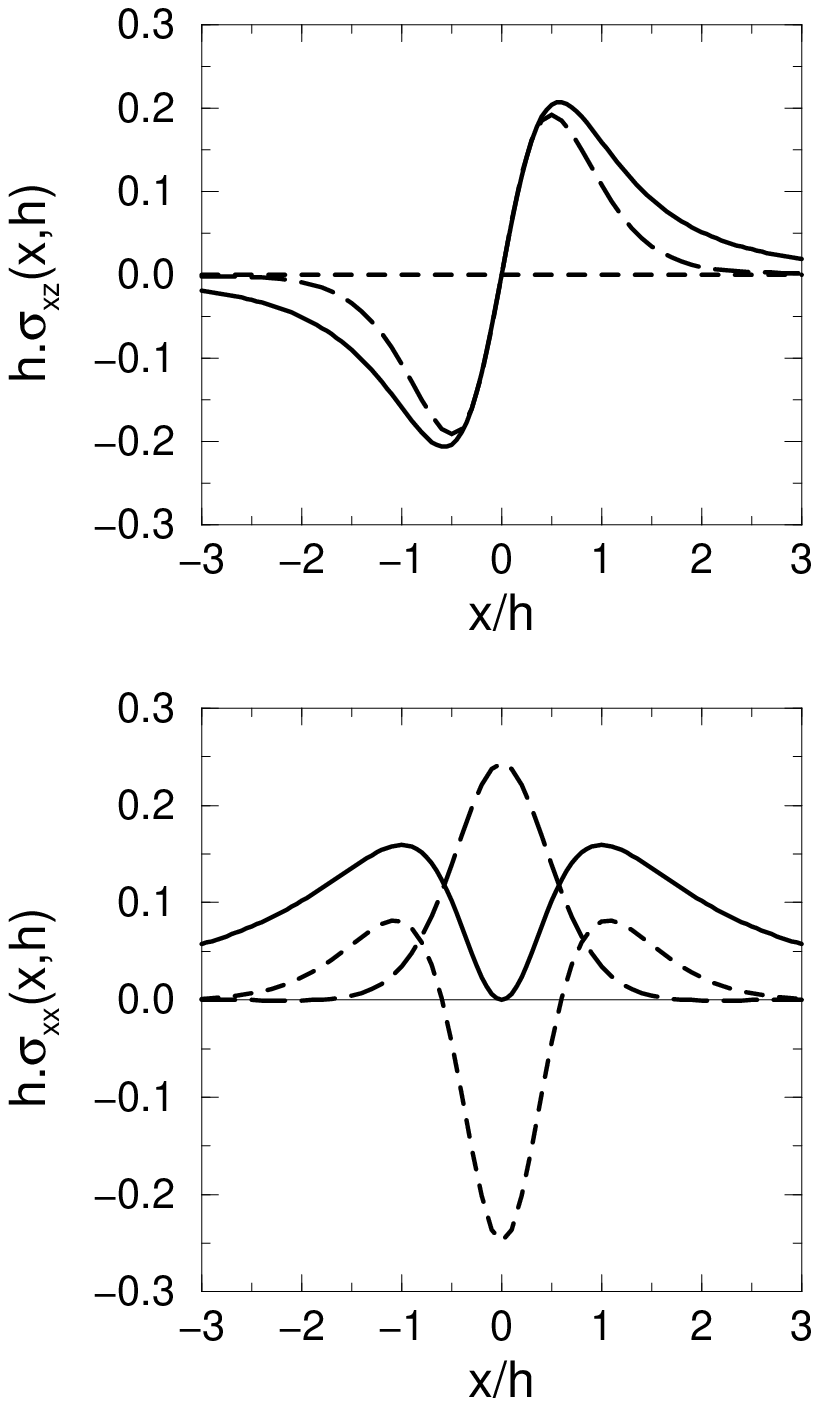}
\caption{\small Stress response functions for a two-dimensional elastic
material. The main plot compares the pressure profile of a semi-infinite
system at depth $z=h$, with the response of a finite elastic layer of
thickness $h$ with either a rough or a smooth bottom. The first and third
curves of each plot are independent of the Poisson coefficient $\nu$. For the
second one, we chose $\nu=0.3$ but its shape depends only very weakly on the
value of $\nu$ -- see inset where the maximum of the response has been plotted
against $\nu$. The side plots show the other components of the stress tensor
(shear and horizontal pressure).
\label{repelastic2d}}
\ec
\efig

In elliptic problems like elasticity, stress or strain conditions must be
specified on \emph{all} the boundaries. Our aim here is to calculate the
response of a layer of height $h$ submitted to a localized pressure at its
top surface. We then suppose that the `piston' which applies this overload
is perfectly smooth and imposes, for example, a normalized ($F=1$) gaussian
profile $Q(x)$ for the vertical pressure 
\begin{equation}
\label{Qdex2dbis}
Q(x) = \frac{1}{\sqrt{2\pi\sigma^2}} \, e^{-x^2/2\sigma^2},
\end{equation}
where $\sigma$ is the adjustable width of this overload. The two conditions
at the top are then (i) $\sigma_{zz}(x,0)=Q(x)$ and (ii) $\sigma_{xz}(x,0)=0$.
Concerning the bottom, we assume that it is perfectly rigid, such that
(iii) $u_z(x,h)=0$, and either very smooth or very rough. The last boundary
condition is then (iv-a) $\sigma_{xz}(x,h)=0$ or (iv-b) $u_x(x,h)=0$
respectively.

\bfig[t]
\bc
\epsfysize=9cm
\epsfbox{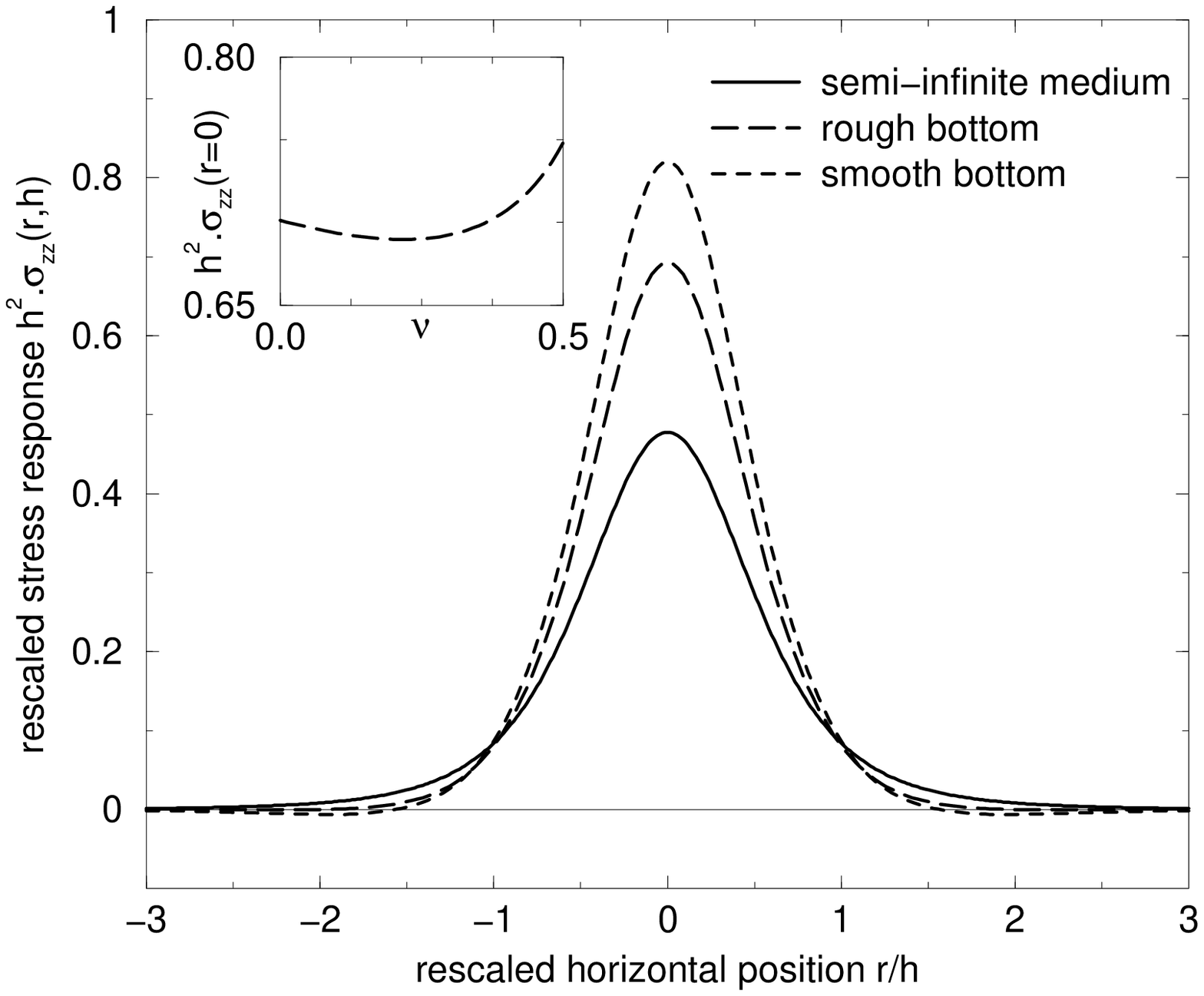}
\hfill
\epsfysize=9cm
\epsfbox{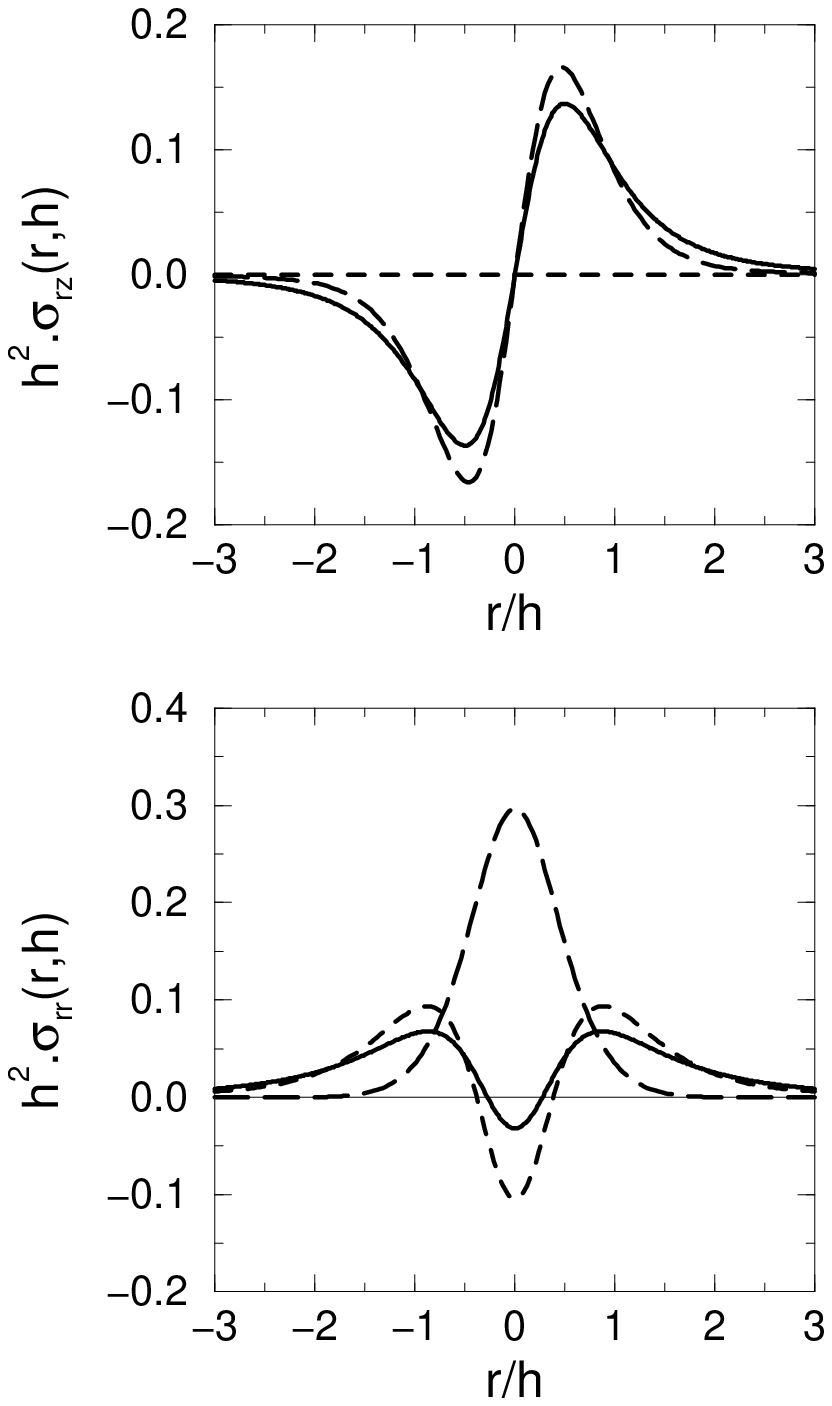}
\caption{\small $3d$ equivalent of figure \protect\ref{repelastic2d}. We chose
again $\sigma=0.001h$ and $\nu=0.3$. The results are qualitatively the same as
in two dimensions.
\label{repelastic3d}}
\ec
\efig

Integrations in both smooth and rough bottom cases can be done numerically,
and the corresponding results for the stress response $\sigma_{ij}$ is
plotted on figure \ref{repelastic2d}. These integrations have been done for
the specific choice of a very peaked gaussian overload: $\sigma=0.001h$ -- a
quasi $\delta$-function. For comparison, these plots are shown together with
the Green's function of a vertically semi-infinite medium.

All three $\sigma_{zz}$ curves have rougly the same shape. This is not
longer true when we plot the horizontal pressure instead of the vertical
one: $\sigma_{xx}$ is proportionnal to $\sigma_{zz}$ (see equation
(\ref{condivb2d})) for a rough bottom, but shows a double peak for a
semi-infinite medium as well as for a smooth bottom -- with a large negative
central part. Negative values can be also seen for $\sigma_{zz}$ on figure 
\ref{repelastic2d} especially for the case of a smooth bottom. They are
absolutely admissible for elastic material -- no delamination between the
material and the bottom is allowed.

An interesting and rather non-intuitive point is that the finiteness of the
elastic layer narrows the stress response. Only the response on the rough
bottom depends on the value of the Poisson coefficient. We chose $\nu=0.3$.
This dependence is very weak for $\sigma_{zz}$ -- see inset of figure
\ref{repelastic2d}. At last, it should be noted that all these curves scale
with the height $h$.

The axi-symmetric three-dimensional calculation is very similar, except that
trigonometric functions have to be replaced by Bessel ones in equations like
(\ref{szzform}). Again, a numerical integration of the functions $A$'s and $B$'s
can be done for a gaussian overload, and the corresponding pressure profiles in
both smooth and rough cases are plotted on figure \ref{repelastic3d}. They are
also compared to the semi-infinite solution. The $3d$ results are qualitatively
the same as in two dimensions. The wideness of the reponse function is non
monotonic with the poisson ratio and presents a maximum for $\nu \sim 0.27$.
A slight difference is that not only the $3d$ solution for the rough bottom
depends on the Poisson coefficient, but the smooth bottom solution too. Again,
this dependence is very weak for the vertical component of the stress tensor.

\section{A quantitative comparison}
\label{fits}

In this section, we want to compare quantitatively the pressure response
measurements with the elastic predictions. Among the two cases calculated in
the previous section (rough and smooth bottom), the first one is the closest
to our experimental situation -- we checked that the shear stress at the
bottom of the grain layer is finite. Therefore, only rough bottom elastic
formulae are going to be used for the following fits.

A set of experimental data is a file with three columns: the horizontal
distance between the piston and the probe $r_k$, the corresponding pressure
measurement $P_k$ and its typical dispersion $\Delta P_k$. There are $N_e
\sim 15$ such triplets for one pressure profile. As explained in section
\ref{exp}, the data have been renormalized by their factor $F^*$ in order to be
of integral unity.

We quantify the `distance' between the experimental pressure profile $P$ and
the elastic predictions $\sigma _{zz}$ by computing the average quadratic
deviation $E$:
\be
\label{ecart}
E^2 = \frac{1}{N_e} \, \sum_{k=1}^{N_e} \left (
\frac{P_k - \szz(r_k,h)}{\Delta P_k} \right )^2.
\ee
$E=1$ would mean that a typical distance between theory and experimental
data is one error bar. A value of $E$ larger than $1$ will then be considered as
not good. Because $\sigma _{zz}$ depends on the Poisson coefficient $\nu $,
$E$ is also a function of $\nu $. This function has a minimal value which gives
the best fitting $\nu $. The precision of this value depends on the sharpness
of this minimum.

\bfig[p]
\bc
\epsfxsize=9cm
\epsfbox{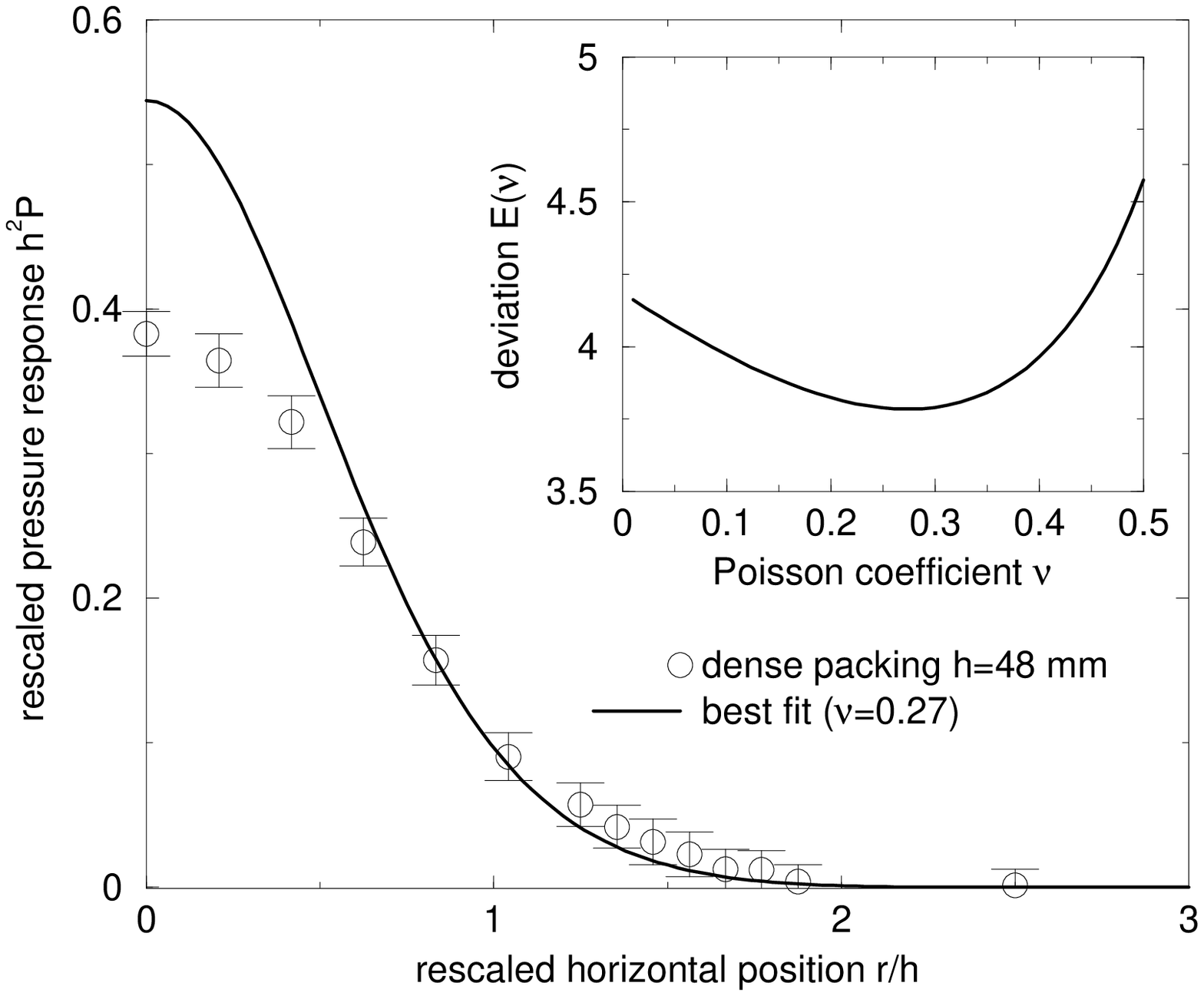}
\caption{\small Fit of the data obtained on a dense packed granular layer. It
is rather poor because the elastic response cannot get wide enough. the inset
shows the deviation $E$ versus $\nu$. $\nu=0.27$ corresponds to $E=3.8$.
\label{fitdense}}
\ec
\efig
\bfig[p]
\bc
\epsfxsize=9cm
\epsfbox{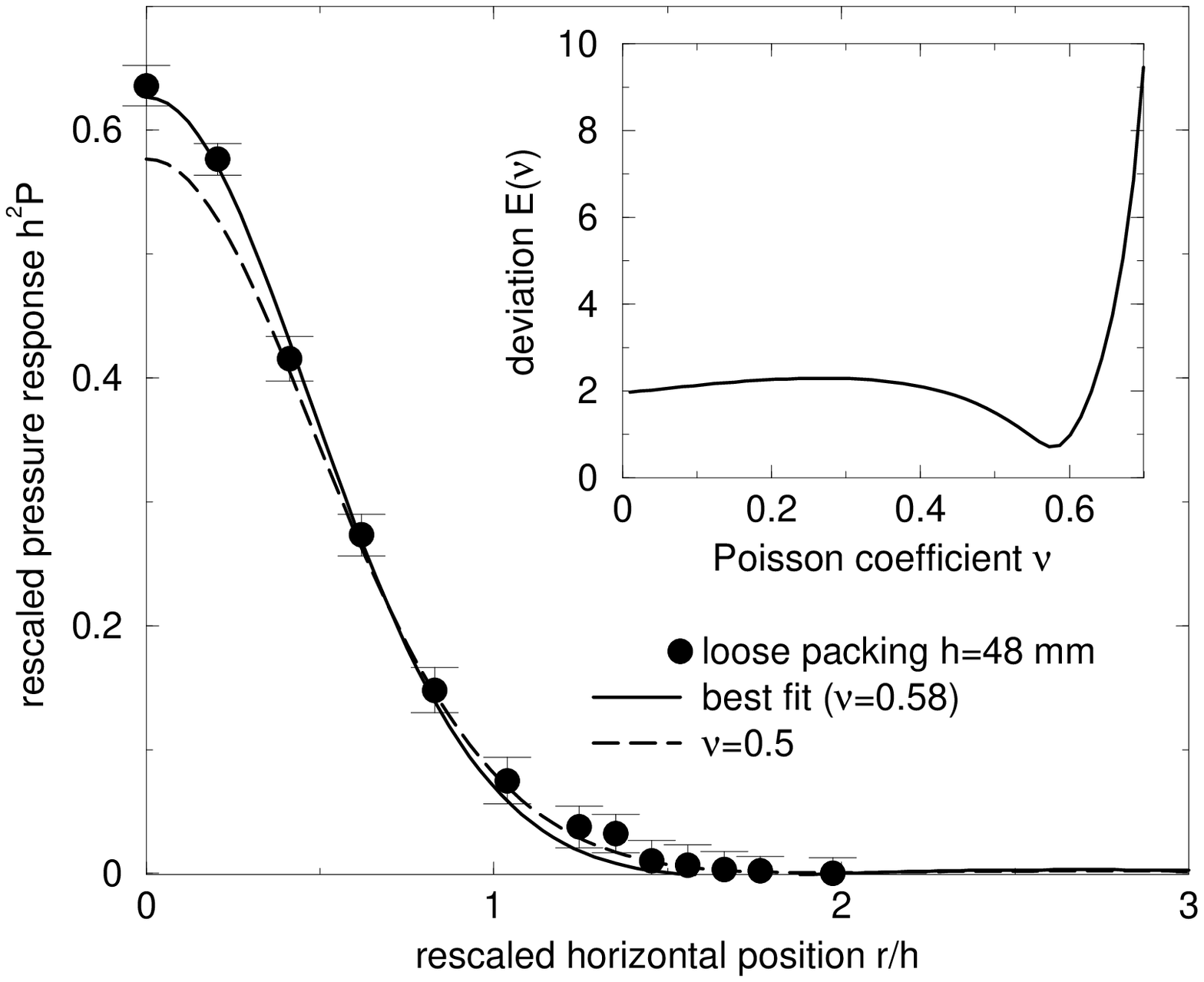}
\caption{\small Fit of the loose packing data. The best Poisson coefficient
exceeds the usual $\nu=\frac{1}{2}$ limit. $\nu=0.58$ corresponds to $E=0.6$,
but $E=1.5$ when $\nu=0.50$.
\label{fitlache}}
\ec
\efig

\bfig[t]
\bc
\epsfxsize=9cm
\epsfbox{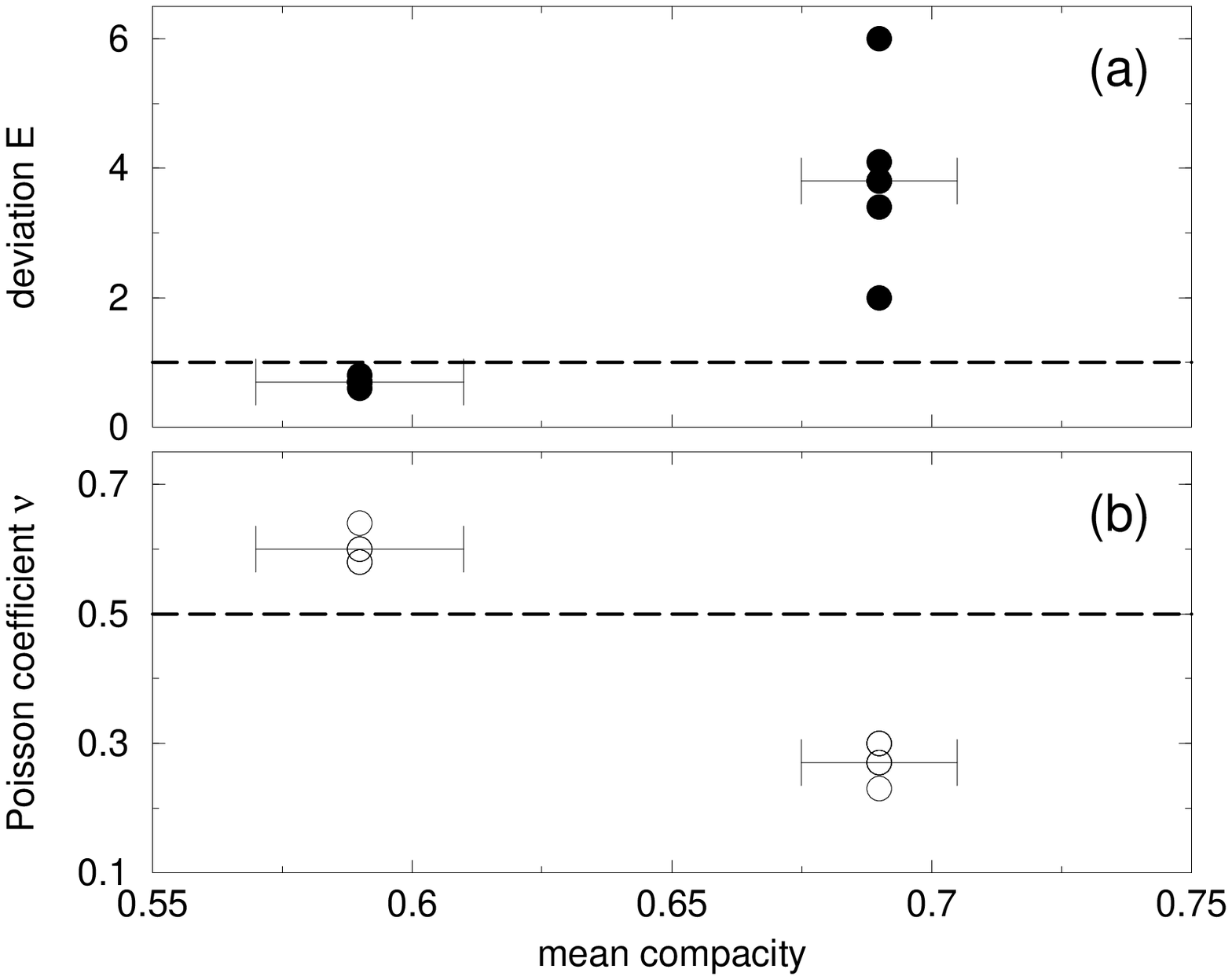}
\caption{\small This figure shows (a) the best elasticity fit quality $E$ and
(b) the corresponding value for the best poisson ration $\nu$ as a function of
the compacity of the packing. Above the dash line $E=1$, the fits are not in
good agreement with the data. Only point under the dash line $\nu=\frac{1}{2}$
are in principle valid in the elastic framework.
\label{allresults}}
\ec
\efig

The results of our fits are gathered together in figure \ref{allresults}.
The results can be summarized as follows. In the case of a dense packing, the
experimental response function is too wide to be well fitted by an elastic
curve -- see figure \ref{fitdense}. As a matter of fact, we get typically
$E\sim 4$ for the best fit. The corresponding Poisson coefficient value does
not then have any real meaning. For the loose packing, the situation is
different. The experimental data, though closer to the elastic response, lay on
a curve which is too narrow to be properly fitted. The best value of $\nu $ is
then $\nu =\frac{1}{2}$ which gives the most narrow elastic response,
corresponding to $E\sim 2$. Interestingly however, if one allows $\nu $ to
exceed the standard limit $\nu =\frac{1}{2}$, one can fit the data pretty
well -- see figure \ref{fitlache}. This exceeding can be done mathematically
because the qualitative shape of the stress profiles calculated with the
elasticity theory changes for $\nu \geq \frac{3}{4}$ only, leading beyond
this value to oscillatory behaviours -- see appendices. Note that the
dilatancy effect in granular material has sometimes been argued to be
somehow encoded by a Poisson coefficient larger than $\frac{1}{2}$, which is
the `incompressibility' limit. Although we do not think that dilatance can
be treated with the concepts of reversible elasticity, our results would
contradict such an argument because only dense packings dilate, loose ones on
the contrary contract.

As we said in section \ref{exp}, the piston which applies the overload at
the top surface of the layer, as well as the pressure probe have an area of
$\sim 1$ cm$^{2}$. Taking into account the finite size of the piston for the
fits was easy in our elastic formalism. Indeed, we assumed that the overload
had a gaussian profile of adjustable width $\sigma $, which was then simply
set to the piston diameter. By contrast, taking into account that of the
pressure probe requires the convolution of the elastic formulae with a disk
of finite diameter. This calculation is much less easy and makes the
computation of the stress profiles difficult to do. In fact, we checked on
few data sets that these two finite size effects are not very important as
soon as the layer thickness $h$ is larger than $\sim 30$ mm, i.e. 3
piston/probe diameters. The actual values of $E$ and $\nu $ that come out
from these modified fits are a bit different -- slightly better -- than that
of figure \ref{allresults}, but the conclusions written in the previous
paragraph keep exactly the same.

\section{Discussion and conclusions}
\label{conclu}

As it was shown that the stress profile under a sandpile does or does not
has a `dip' below the apex of the pile, we found that the pressure response
of a layer of sand submitted to a localized normal force at its top surface
depends on the way this layer has been built -- its `history'. The response
is rather wide when the grain packing is made very dense and compact, but it
is more narrow when the layer is loose. For a given height, the maximal
value of the pressure is approximately twice smaller in the first than in
the second case. The predictions of isotropic elasticity theory, even when
taking into account the finite layer thickness, agree poorly with the
experimental data.  However note the puzzling result that, taking a Poisson
coefficient larger than the usual limit $\frac{1}{2},$ can fit rather well
the experimental data for the loose packing preparation. We have no
interpretation of this fact besides concluding for the non adequacy of the
isotropic elastic picture for piling prepared using  `dense' or `loose'
filling procedures.

Although we present all our results in terms of dense or loose packing, the
compacity in itself is certainly not a good control parameter. Rather, a
natural way to improve the fits within this elastic framework would be to
take into account a possible anisotropy of the material. A classical example
is the so-called `aelotropy' which occurs when the vertical symmetry axis
has different mechanical properties than the horizontal directions. Such an
anisotropy has five independent parameters: two Poisson coefficients (a
vertical and a horizontal one), two Young moduli (idem) and a shear modulus.
A theory with so many parameters will for sure fit our data. Indeed, the
shear modulus has been found to be of strong influence on the shape of the
response function \cite{Garnier}.

We think however that a standard elastic description of granular materials
is unsatisfactory: a proper definition of the kinetic variables may be
problematic for systems of hard particles. As a matter of fact, the
link between the local microscopic movement of the grains and the possible
corresponding large scale displacement field is a current subject of
research \cite{Cambou}. Recent numerical simulations of frictionless disks
even suggest that the stress-strain relation might not converge to a well
defined curve for larger and larger systems \cite{Combe}.

A rather striking feature of granular systems is the presence of force
chains. These chains support most of the weight of the grains, and their
geometrical characteristics -- length, orientation -- is the signature of the
history of the system. In a recent paper \cite{YY}, some of us with others have
shown that it is possible to get pseudo-elastic equations from a simple model of
-- perfectly rigid -- force chains which can split or merge at some
`defects' of the grain packing. In this model however, the stress tensor 
\emph{as well as} the vector field which plays the role of the displacement
in elasticity can be both built from the angular distribution of force chains.
The specification of the boundary conditions is therefore a non trivial issue
on which we are currently working. There are two main advantages in this new
approach. First, no real displacement field is needed, and second, it allows to
calculate the pseudo-elastic coefficient from microscopic quantities -- the force
chains angular distribution. The idea is then to introduce some anisotropy in this
distribution, and see which kind of anisotropic pseudo-elastic equations we get
out of it -- note that this work would be very close in spirit to e.g.
\cite{Calvetti,Radjai2} were they try to link local geometrical variables
(orientation of contacts between grains) with the mechanical properties at larger
scales. The fit of our data would then give an information on the local structure
of the packing. More response function experiments are thus planed to be
performed with new preparation history, in particular with shearing or
avalanching procedures in order to be able to come back to the yet unresolved
sandpile `dip' problem.

\centerline{\rule[0.1cm]{5cm}{1pt}}

It is a pleasure to thank R.P. Behringer, J.-P. Bouchaud, M.E. Cates, J.
Geng, M. Otto, Y. Roichman, D.G. Schaeffer, J.E.S. Socolar and J.P. Wittmer
for very useful discussions on the problem of the response function of a
granular layer. We are grateful to G. Ovarlez who did a careful check of the
elastic calculations, and to E. Flavigny for making us be aware of references
\cite{Giroud} and \cite{Garnier}.
This work has been partially supported by an Aly Kaufman postdoctoral
fellowship. We acknowledge the financial support of the grant PICS-CNRS 563.
D.L. acknowledges support from U.S. - Israel Binational Science Foundation 
grant 1999235.


\appendix
\newpage

\section{Appendix: the two-dimensional elastic calculation}
\label{calcul2d}

For a two-dimensional elastic layer, we have three independent stress
tensor components: the pressures $\sxx$ and $\szz$, and the shear $\sxz$. $x$
is the horizontal axis, and $z$ is the vertical one, pointed downwards. The
continuity equations (\ref{equil}) can then be explicitly written down as
follows:
\bea
\label{equiz2d}
\dr_z \szz + \dr_x \sxz & = & \rho g \\
\label{equix2d}
\dr_z \sxz + \dr_x \sxx & = & 0.
\eea
Besides these two equilibrium equations, an additionnal and independent
equation is required to solve the problem. Among equations
(\ref{equasLandau}), the simplest is
\be
\label{laplace}
\De (\szz + \sxx) = 0.
\ee
It is natural in this context to introduce the new variables
\bea
\label{defT2d}
T & = & \szz + \sxx\\
\label{deftau2d}
\tau & = & \sxz\\
\label{defD2d}
D & = & \szz - \sxx.
\eea
Using equations (\ref{equiz2d}) and (\ref{equix2d}), it is easy to show that
these new functions verify
\bea
\label{equaT2d}
\De T & = & 0\\
\label{equatau2d}
\De \tau & = & - \, \dr_x \dr_z T\\
\label{equaD2d}
\dr_x \dr_z D & = & (\dr_z^2 - \dr_x^2) \tau.
\eea

A standard mathematical base of harmonic functions is the product of
trigonometric functions with exponentials. We shall keep to
$x \leftrightarrow -x$ symmetrical situations such that we can look
for a solution of the type
\bea
\label{solT2d}
T & = & B_1z + C_1
    + \intq \cos(qx) \left [ a(q) e^{qz} + b(q) e^{-qz} \right ] \\
\label{soltau2d}
\tau & = & \frac{1}{2} \, B_2x
       + \intq \sin(qx)
         \left \{
         \left [ c(q) e^{qz} + d(q) e^{-qz} \right ] +
         \frac{1}{2} \, qz
         \left [ a(q) e^{qz} + b(q) e^{-qz} \right ]
         \right \} \\
\nonumber
D & = & 2 \rho g z -(B_1+B_2)z + C_2 \\
\label{solD2d}
  &   & - \intq \cos(qx)
          \left \{
          qz \left [ a(q) e^{qz} - b(q) e^{-qz} \right ] +
          2  \left [ c(q) e^{qz} - d(q) e^{-qz} \right ]
          \right \}
\eea
where the constants $B_1$, $B_2$, $C_1$ and $C_2$, as well as the the
functions $a(q)$, $b(q)$, $c(q)$ and $d(q)$ are to be determined by the
boundary conditions.

\subsection*{Boundary conditions}

We suppose that the top surface is submitted to a localized vertical and
unitary overload, for example a normalized gaussian profile $Q(x)$ for the
vertical pressure:
\be
\label{Qdex2d}
Q(x) = \frac{1}{\sqrt{2\pi\sigma^2}} \, e^{-x^2/2\sigma^2}
\ee
$\sigma$ is the adjustable width of this overload.
The two conditions at the top are then (i) $\szz(x,0)=Q(x)$ and (ii)
$\sxz(x,0)=0$. Concerning the bottom, we assume that it is perfectly
rigid, such that (iii) $u_z(x,h)=0$, and either very smooth or very
rough. The last boundary condition is then (iv-a) $\sxz(x,h)=0$ or
(iv-b) $u_x(x,h)=0$ respectively.

It is convenient for our calculation to transform the displacement
conditions (iii) and (iv-b) into stress conditions. For that purpose,
one can take derivatives of condition (iii) with respect to $x$ and
introduce stress components via the strain-stress relations. Using
also equilibrium equations (\ref{equiz2d}) and (\ref{equix2d}), one finally
gets for the condition (iii):
\be
\label{condiii2d}
(2+\nu) \dr_x \sxz(x,h)= \dr_z \sxx(x,h) - \nu \rho g.
\ee
A similar calculation leads to the following new condition (iv-b):
\be
\label{condivb2d}
\sxx(x,h) = \nu \szz(x,h).
\ee

Since we look for a solution of the form of (\ref{solT2d}), (\ref{soltau2d})
and (\ref{solD2d}), it is natural to introduce the function $s(q)$ such that
\be
\label{Qdes2d}
Q(x) = \intq \cos(qx) s(q),
\ee
which, for the specific gaussian choice (\ref{Qdex2d}) leads to
\be
\label{sdeq2d}
s(q) = \frac{1}{\pi} \, e^{-\sigma^2q^2/2}.
\ee

\subsection*{Solution for a smooth bottom}

We switch gravity off since it is not of interest for the calculation of
the response function. A simple term to term identification in boundary
conditions (i), (ii), (iii) and (iv-a) leads to vanishing coefficients
$B_i$ and $C_i$, and to the four following linear equations for the
unknown functions $a(q)$, $b(q)$, $c(q)$ and $d(q)$:
\bea
\label{condI2d}
\frac{1}{2}[a(q) + b(q)] - c(q) + d(q) & = & s(q)\\
\label{condII2d}
c(q) + d(q) & = & 0\\
\label{condIII2d}
(1+\nu) \left \{
\left [ c(q) + \frac{1}{2} qh \, a(q) \right ] e^{qh} +
\left [ d(q) + \frac{1}{2} qh \, b(q) \right ] e^{-qh}
\right \}
& = & a(q) e^{qh} - b(q) e^{-qh}\\
\label{condIVa2d}
\left [ c(q) + \frac{1}{2} qh \, a(q) \right ] e^{qh} +
\left [ d(q) + \frac{1}{2} qh \, b(q) \right ] e^{-qh} & = & 0,
\eea
whose solution is
\bea
\label{asmooth2d}
a(q) & = & 2 s(q) \, \frac{\sinh(qh) e^{-qh}}{\sinh(2qh) + 2qh}\\
\label{bsmooth2d}
b(q) & = & 2 s(q) \, \frac{\sinh(qh) e^{qh}}{\sinh(2qh) + 2qh}\\
\label{cetdsmooth2d}
c(q) & = & - d(q) = -\frac{1}{2} \, s(q) \,
\frac {2qh}{\sinh(2qh) + 2qh}.
\eea
Putting these relations back to the equations (\ref{solT2d}),
(\ref{soltau2d}) and (\ref{solD2d}), we get explicit -- integral --
expressions for the stress components. Note that the function $a$,
$b$, $c$ and $d$ are independent of $\nu$.

\subsection*{Solution for a rough bottom}

For the case of a rough bottom, the condition (iv-a) has to be
replaced by the condition (iv-b). It means that the last equation of the
system (\ref{condI2d}-\ref{condIVa2d}) has to be changed into
\be
\label{condIVb2d}
- \left [ c(q) + \frac{1}{2} qh \, a(q) \right ] e^{qh}
+ \left [ d(q) + \frac{1}{2} qh \, b(q) \right ] e^{-qh}
= \frac{1}{2} \, \frac{1-\nu}{1+\nu}
\left [ a(q) e^{qh} + b(q) e^{-qh} \right ]
\ee
and the resolution of these four linear equations leads this time to the
following solution
\bea
\label{arough2d}
a(q) & = & 2 s(q) \,
\frac{f_-(q) + 2qh}{f_+(q)f_-(q) + 4q^2h^2}\\
\label{brough2d}
b(q) & = & 2 s(q) \,
\frac{f_+(q) - 2qh}{f_+(q)f_-(q) + 4q^2h^2}\\
\label{cetdrough2d}
c(q) & = & -d(q) = -\frac{1}{2} \, s(q) \,
\left [ 1 - 
\frac {f_+(q) + f_-(q)}{f_+(q)f_-(q) + 4q^2h^2}
\right ],
\eea
where the functions $f_+$ and $f_-$ are defined by
\be
\label{fpm2d}
f_{\pm}(q) = 1 + \frac{3-\nu}{1+\nu} e^{\pm 2qh}.
\ee
This time, there is a dependance in $\nu$. In principle, the Poisson
coefficient should be less that unity in 2$d$. However, these functions
really change behaviour only for $\nu \ge 3$, leading to oscillatory
stresses -- they however develop negative parts as $\nu \to 3$.

\subsection*{Semi-infinite medium}

For comparison, in the case of a semi-infinite medium submitted to a
ponctual and vertical unitary force at $x=0$, the stress components
are given \cite{Landau} by
\bea
\label{szzsi2d}
\szz & = & \frac{2}{\pi} \, \frac{z^3}{(x^2 + z^2)^2}   \\
\label{sxxsi2d}
\sxx & = & \frac{2}{\pi} \, \frac{z x^2}{(x^2 + z^2)^2} \\
\label{sxzsi2d}
\sxz & = & \frac{2}{\pi} \, \frac{x z^2}{(x^2 + z^2)^2}.
\eea

\section{Appendix: the three-dimensional case}
\label{calcul3d}

The calculation in the three-dimensional case is very similar to the 2$d$ one.
We shall keep to axi-symmetric situations so that the stress tensor has only
four non-zero components: the pressures $\szz$, $\srr$, $\stt$ and the shear
$\srz$. $z$ is again the vertical axis pointing downwards, and $(r,\theta)$
are the horizontal planar coordinates.

The equations we want to solve are simpler with the new functions
\bea
\label{defT3d}
T & = & \szz + \srr + \stt \\
\label{deftau3d}
\tau & = & \srz \\
\label{defS3d}
S & = & \srr + \stt \\
\label{defD3d}
D & = & \srr - \stt,
\eea
which must verify
\bea
\label{equaT3d}
\De T & = & 0 \\
\label{equaS3d}
(1+\nu) \De S & = & \dr_z^2 T \\
\label{equatau3d}
\dr_r \tau + \frac{\tau}{r} & = & - \dr_z (T-S) + \rho g \\
\label{equaD3d}
\dr_r D + 2\frac{D}{r} & = & -\dr_r S - 2\dr_z \tau.
\eea
The two last equations are the explicit forms of the force balance equations
(\ref{equil}), and we got the two first ones from equations
(\ref{equasLandau}).

The corresponding general solutions involve Bessel functions of the first kind
$J_0$, $J_1$ and $J_2$, and read:
\bea
\label{solT3d}
T & = & T_1z + T_2
    + \intq J_0(qr) \left [ a(q) e^{qz} + b(q) e^{-qz} \right ] \\
\nonumber
S & = & S_1z + S_2 \\
\label{solS3d}
 & & + \intq J_0(qr) \left \{
       \left [ c(q) e^{qz} + d(q) e^{-qz} \right ] +
       \frac{1}{2} \, \frac{1}{1+\nu}  \, qz
       \left [ a(q) e^{qz} - b(q) e^{-qz} \right ]
       \right  \} \\
\nonumber
\tau & = & \frac{1}{2} \, \rho gr + \frac{1}{2} \, r(S_1-T_1) + \frac{u(z)}{r}
       + \intq J_1(qr)
         \left [ [c(q)-a(q)] e^{qz} + [d(q)-b(q)] e^{-qz} \right ] \\
\label{soltau3d}
 & &   + \frac{1}{2} \, \frac{1}{1+\nu} \intq J_1(qr)
         \left [ a(q)[1+qz] e^{qz} - b(q)[1-qz] e^{-qz} \right ] \\
\nonumber
D & = & \frac{v(z)}{r^2} - \frac{du(z)}{dz}
    + \intq J_2(qr)
      \left [ [2a(q)-c(q)] e^{qz} + [2b(q)-d(q)] e^{-qz} \right ] \\
\label{solD3d}
 & & - \frac{1}{1+\nu} \intq J_2(qr)
       \left [ a(q)[2+\frac{1}{2} \,qz] e^{qz}
             + b(q)[2-\frac{1}{2} \,qz] e^{-qz} \right ],
\eea
where the constants $T_1$, $T_2$, $S_1$ and $S_2$, as well as the functions 
$u(z)$, $v(z)$, $a(q)$, $b(q)$, $c(q)$ and $d(q)$ are, again, to be determined
by the boundary conditions.

\subsection*{Boundary conditions}

As in the 2$d$ case, we want to impose at the surface (i) an overload
$\szz(r,0)=Q(r)$, but (ii) no shear $\srz(r,0)=0$. At the bottom, the
vertical displacement must vanish $\uz(r,h)=0$ (iii), and we shall study the
two cases, very smooth $\srz(r,h)=0$ (iv-a) or very rough $\ur(r,h)=0$ (iv-b)
bottom.

The 3$d$ equivalent of (\ref{Qdex2d}) is now
\be
\label{Qder3d}
Q(r) = \frac{1}{2\pi\sigma^2} \, e^{-r^2/2\sigma^2}.
\ee
Looking at the general form of the solution (\ref{solT3d}-\ref{solD3d}), it is
natural to introduce the function $s(q)$ defined by the relation
\be
\label{Qdes3d}
Q(r) = \intq J_0(qr) s(q)
\ee
which, for $Q$ given by (\ref{Qder3d}) gives
\be
\label{sdeq3d}
s(q) = \frac{1}{2\pi} \, q \, e^{-\sigma^2q^2/2}.
\ee

Taking derivatives of the conditions on displacements and using stress-strain
relations, one can again transform these conditions into relations between
stress components only. It is easy to show that the condition (iii) can be
written as
\be
\label{condiii3d}
2(1+\nu) \dr_r \tau = \frac{1}{2} (1+\nu) \dr_z (S+D) - \nu \dr_z T,
\ee
and that the condition (iv-b) gives
\be
\label{condivb3d}
(1+\nu)(S+D) = 2\nu T
\ee
(these last two relations are only valid at $z=h$).

\subsection*{Solution for a smooth bottom}

Again, as we are interested to the response of the elastic layer to
this overload, gravity is switched off. The four conditions (i)-(iv-a) then
give four equations for the unknown functions $a(q)$, $b(q)$, $c(q)$ and
$d(q)$, all other functions and constants being zero. These equations are:
\bea
\label{condI3d}
a(q)+b(q)-c(q)-d(q) & = & s(q) \\
\label{condII3d}
c(q)-d(q) & = & \frac{1+2\nu}{2(1+\nu)} \, [a(q)-b(q)] \\
\label{condIII3d}
2(1+\nu) \left [ c(q) e^{qh} - d(q) e^{-qh} \right ] & = &
(3-qh) a(q) e^{qh} - (3+qh) b(q) e^{-qh} \\
\label{condIVa3d}
2(1+\nu) \left [ c(q) e^{qh} - d(q) e^{-qh} \right ] & = &
(1+2\nu-qh) a(q) e^{qh} - (1+2\nu+qh) b(q) e^{-qh},
\eea
whose solution is
\bea
\label{asmooth3d}
a(q) & = & 2s(q) (1+\nu) \frac{\sinh(qh)e^{-qh}}{\sinh(2qh)+2qh} \\
\label{bsmooth3d}
b(q) & = & 2s(q) (1+\nu) \frac{\sinh(qh)e^{qh}}{\sinh(2qh)+2qh} \\
\label{csmooth3d}
c(q) & = & -\frac{1}{2} \, s(q) \left [
1 - (3+4\nu)\frac{\sinh(qh)e^{-qh}}{\sinh(2qh)+2qh}
  -         \frac{\sinh(qh)e^{qh}}{\sinh(2qh)+2qh} \right ] \\
\label{dsmooth3d}
d(q) & = & -\frac{1}{2} \, s(q) \left [
1 -         \frac{\sinh(qh)e^{-qh}}{\sinh(2qh)+2qh}
  - (3+4\nu)\frac{\sinh(qh)e^{qh}}{\sinh(2qh)+2qh} \right ]
\eea

\subsection*{Solution for a rough bottom}

In the case of a very rough bottom, the relation (\ref{condIVa3d}) should be
replaced by the condition (iv-b), i.e. by:
\be
\label{condIVb3d}
2(1+\nu) \left [ c(q) e^{qh} + d(q) e^{-qh} \right ] = 
(4\nu-qh) a(q) e^{qh} + (4\nu+qh) b(q) e^{-qh}.
\ee
The resolution of the four equations then gives
\bea
\label{arough3d}
a(q) & = & 2s(q) (1+\nu) \frac{f_-(q)+2qh}{f_+(q)f_-(q)+4q^2h^2} \\
\label{brough3d}
b(q) & = & 2s(q) (1+\nu) \frac{f_+(q)-2qh}{f_+(q)f_-(q)+4q^2h^2} \\
\label{crough3d}
c(q) & = & -\frac{1}{2} \, s(q)
\left [
1 - \frac{(3+4\nu)f_-(q)+f_+(q)+4qh(1+2\nu)}{f_+(q)f_-(q)+4q^2h^2}
\right ] \\
\label{drough3d}
d(q) & = & -\frac{1}{2} \, s(q)
\left [
1 - \frac{(3+4\nu)f_+(q)+f_-(q)-4qh(1+2\nu)}{f_+(q)f_-(q)+4q^2h^2},
\right ]
\eea
where the functions $f_+$ and $f_-$ are defined by
\be
\label{deffpm3d}
f_\pm (q) = 1 + (3-4\nu) \, e^{\pm 2qh}.
\ee
Again, as in the $2d$ case, the standard limit $\nu=\frac{1}{2}$ can
be exceeded without quantitative change -- except the appearance of negative
parts --, up to $\nu=\frac{3}{4}$ where the stresses start oscillating.

\subsection*{Semi-infinite medium}

For comparison, in the case of a semi-infinite medium submitted to a
ponctual and vertical unitary force at $r=0$, the stress components
are given by Boussinesq and Cerruti's formulae \cite{Johnson}:
\bea
\label{szzsi3d}
\szz & = & \frac{3}{2\pi} \, \frac{z^3}{(r^2 + z^2)^{5/2}} \\
\label{srrsi3d} \srr & = & \frac{1}{2\pi}
\left [
(1-2\nu) \left ( \frac{1}{r^2} - \frac{z}{r^2(r^2 + z^2)^{1/2}} \right )
-\frac{3 z r^2}{(r^2 + z^2)^{5/2}}
\right ] \\
\label{sttsi3d}
\stt & = & \frac{1}{2\pi} (1-2\nu)
\left (
\frac{1}{r^2} - \frac{z}{r^2(r^2 + z^2)^{1/2}} - \frac{z}{(r^2 + z^2)^{3/2}}
\right ) \\
\label{srzsi3d}
\srz & = & \frac{3}{2\pi} \, \frac{r z^2}{(r^2 + z^2)^{5/2}}.
\eea

\end{document}